\documentclass[11pt]{article} 
\usepackage{rldmsubmit,palatino}
\usepackage{graphicx}
\usepackage{hyperref}

\usepackage[
backend=biber,
]{biblatex}
\title{A Human-Centered Approach \\ to Interactive Machine Learning}

\author{
Kory W.~Mathewson\thanks{https://korymathewson.com} \\
Department of Computing Science \\
University of Alberta \\
Edmonton, Alberta, Canada \\
\texttt{korymath@gmail.com}
}

\begin{document}

\maketitle

\begin{abstract}
The interactive machine learning (IML) community aims to augment humans' ability to learn and make decisions over time through the development of automated decision-making systems. This interaction represents a collaboration between multiple intelligent systems---humans and machines. A lack of appropriate consideration for the humans involved can lead to problematic system behaviour, and issues of fairness, accountability, and transparency. This work presents a human-centred thinking approach to applying IML methods. This guide is intended to be used by AI practitioners who incorporate human factors in their work. These practitioners are responsible for the health, safety, and well-being of interacting humans. An obligation of responsibility for public interaction means acting with integrity, honesty, fairness, and abiding by applicable legal statutes. With these values and principles in mind, we as a research community can better achieve the collective goal of augmenting human ability. This practical guide aims to support many of the responsible decisions necessary throughout iterative design, development, and dissemination of IML systems.
\end{abstract}


\acknowledgements{This research was partially supported by the University of Alberta, Alberta Machine Intelligence Institute (Amii), the Natural Sciences and Engineering Research Council (NSERC), and the Alberta Innovates Technology Fund (AITF). I am grateful to Dr. Patrick Pilarski, Dr. Matt Taylor, Alex Kearney, Johannes G\"{u}nther, Jamie Brew, Lana Cuthbertson, Katya Kudashkina, Keyfer Mathewson, and Dr. Piotr Mirowski for sharing their comments during the preparation of this manuscript. Any errors are my own.}

\startmain

\section{Introduction and Background}
\label{sec:introbg}

Machine learning (ML) comprises a set of computing science techniques for automating knowledge acquisition rather than relying on explicit human instruction. Interactive ML (IML) systems incorporate humans with ML techniques. The goals of IML include the amplification and augmentation of human abilities. All ML system have humans at some point in the learning loop; some interactions are more evident than others. For example, interacting humans may provide data, objective functions, direct feedback, algorithms, or code. Some system designers choose to acknowledge and feel a responsibility for these humans more than others.

Examining ML from a human-centred perspective requires re-framing work-flows to include humans during conceptualization, implementation, and interaction. Human-centred thinking will lead to more usable ML systems, improved mutual understanding, and augmented communication between intelligent systems \cite{pilarski2017communicative}. These benefits come at the increased cost of time and energy to understand human factors and many designers choose to ignore social factors. Ignorance has serious implications (i.e. reward misalignment between humans and machines \cite{fatml, char2018implementing, zafar2017fairness,leike2018scalable}.

Human-centred design develops solutions to problems by involving the human perspective in all steps of the process. Human-centred systems put people before machines, and delight in the ability and ingenuity of humans \cite{cooley1996human}.

Fails and Olsen (2003) \cite{Fails:2003:IML:604045.604056} introduced interactive machine learning (IML). Amershi \textit{et al.} (2014) \cite{amershi2014power} discuss interaction modalities, and the AAAI-17 tutorial, ``Interactive Machine Learning: From Classifiers to Robotics'', presented a comprehensive history of the field \cite{aaai2017tut}. Incorporating human decision making with ML systems continues to be an active area of research \cite{DBLP:journals/corr/MathewsonP16,DBLP:journals/corr/abs-1811-06521,leike2018scalable,DBLP:journals/corr/abs-1709-03969}. Previous work has discussed the virtues of human-centered thinking in ML \cite{hcml,lovejoy2017human}. But, there continues to be a gap between the mental models of responsible ML practitioners and many contemporary ML deployments.

A good proportion of ML research happens at institutions with ethics review boards. These groups should rigorously apply ethical principles through review of proposed studies. But, they may not exist in your organization, or they may not provide a comprehensive review. This guide details many of the ethics review board questions and extends beyond these considerations. Many ML systems will find their way out of laboratory settings without ethics review. For instance, many ML studies will have no human subjects, but this research (i.e. the code, data, and models) will be released and used by the general public. Design, development, and deployment can happen without subscribing human subjects directly.

Sec. \ref{sec:hcd} focuses on the humans-in-the-loop, the goals and hypotheses of the research, and the data. This section also presents two design thinking exercises: a \textit{whiteboard model}--which can help identify streamlined processes, and a \textit{pre-mortem}--which can help identify potential risks and failure modes early in the process. Sec. \ref{sec:dev} concerns itself with the the iterative development process. It advocates for checking in with the humans-in-the-loop at each iteration. Sec. \ref{sec:diss} focuses on questions of model deployment and public communication. It can be helpful to read through this guide before starting the project and to continually check-in with it for design, development, and dissemination of human-interactive decision-making systems.

\section{Human-Centered Design}
\label{sec:hcd}

\paragraph{Step 1: Define the hypothesis} State the investigated question of interest. Can you pose it as a testable hypothesis \cite{popper2005logic}, which can be supported by evidence? The premise is the starting point and motivation for investigation.

\paragraph{Step 2: Loop in humans} Define your values and principles. How will you align this work with human needs? Identify individual and societal social factors in the problem. Start by considering why your hypothesis is essential not just to you, but to the larger community of people who might interact with it. How will the system augment human lives?

Consider the question of interest from multiple stakeholder perspectives. For instance, think about three groups of individuals: those invested in the success of the work, those impacted by your work, and those who might be interested in the work. Empathize with stakeholders to appreciate how the problem, and potential solutions, will affect them.

Consent starts with communication. Build open-communication channels with stakeholders in these groups. Gather from them ideas, design requirements, concerns, and questions. You should review your values alongside the values of these individuals. How will these stakeholders engage with your system? Thinking about this now will help during Step 9: Deploy. How might the interaction look? Ask each of them how they will evaluate the performance of your system? Discuss how the system might be used both constructively and misapplied to harm. This dual-use discussion is ongoing in the field of ML \cite{gpt2}.

Choices you make will impact these people directly, and you are responsible for the impact of your work on them. Embrace this responsibility. These stakeholders can champion your system if you engage them early in the process and often through iteration.

\paragraph{Step 3: Define the goal} Define a specific, measurable, attainable, relevant, and timely goal. It should be linked directly to the hypothesis from Step 1: Hypothesis. The goal should clearly define success. This definition will embrace the ways that your stakeholders will engage with, and evaluate the performance of your system.

Often there are multiple metrics which define success for a given problem. ML system designers often refine optimization to a single metric of interest. Consider both your optimization metric (i.e. model performance indicator) and your measures of system success. How does this learning objective align with the ways that your stakeholders will evaluate your system? Define a testing suite for safety which you will use in Step 6: Evaluate your model. These tests should consider human health, safety, and well-being. As well, they should evaluate your system on biases, fairness, and equality across hidden features in your data \cite{DBLP:journals/corr/abs-1711-09883}.

It can help to align your work with familiar categories of existing ML work \cite{langley2000crafting}. Is this project developing a new model, applying existing methods to new data, or presenting a new model of human behaviour? Does the system test the limitations of current models on a new problem?

Given your goal, what are the technical, scientific, implementation problems which need to be solved? You should be able to break your system into small components (e.g. data, processing, evaluation). Doing so will make addressing each part individually easier. What are some of the downstream impacts of accomplishing your goal? That is, if the results support your hypothesis, what else might be true?

\paragraph{Step 4: Define the data} An ML system is a reflection of the training data it learns on. It can reflect many common human biases. What is your ideal data set? How much data do you desire? How much data do you need? Why might these amounts be different? What are the dependent and independent variables? How will the data be organized and represented for the learning system? How might possible data sources stray from the ideal ort data? How will you define what outliers, and bad data points, are?

How will you accumulate, clean, parse, label, and safely store your data? How might you fill in blanks in your data? Can you use software to simulate data? Experiments on simulated data designed to test assumptions and gain intuition provide valuable insights. How will you incorporate new data which arrives after deployment?

How will you handle participant recruitment and compensation? If you pay for data (e.g. through crowd-sourcing, direct payment to humans, or a third party), what are the costs of accumulating data? What are the usage rights and responsibilities of your data? What is the ownership model for this data?

If you have humans in your data collection, consider the ethical implications of collecting their data. How are your data generating humans informed of the use of their data? How are data privacy and security communicated? What are the potential biases and sensitivities in human-collected data (i.e. personal or identifying information)?

Once you collect your data, split it into training, evaluation (i.e. validation), and held-out testing data segments. Do this early, lest you leak information from the test segment into your model selection and parameter tuning processes.

\subsection{Design Thinking Exercises}

\paragraph{The Whiteboard Model}
For the whiteboard model exercise consider the following: given your hypothesis, stakeholder analysis, and goal, how would you get to a solution given a short amount of time and only a whiteboard? It is tempting to think about novel techniques which might address your goal. Preferably, it is often more effective to make something that works (i.e. your whiteboard model) and then make iterative improvements. This thought exercise will also provide an opportunity to mentally zoom-out from the problem and think about how potential solutions fit into a broader direction.

\paragraph{The Pre-mortem}
For the pre-mortem exercise, imagine that the project fails for a variety of reasons. Write down these failure modes. Then, for each failure mode, work backwards to identify what might have lead to different results. This process of prospective hindsight can increase the ability to identify reasons for future outcomes by 30\% \cite{mitchell1989back} correctly. A pre-mortem can provide insights and ideas which you can use in the next iterative development steps and can help reduce the chances of arriving at predictable failure modes.

\section{Develop, Analyze, Evaluate, and Iterate}
\label{sec:dev}

\paragraph{Step 5: Build model} Safe design is the first step towards safe use. Think about model misuse starting with the first model you build. Step 2: Loop in Humans covered much of this preparation.

Consider simple models for your learning from your data. A simple model serves as a baseline for comparing model improvements. What is your baseline model? It might be a model that generates random outputs; \textit{random} is a perfectly reasonable baseline and can help to identify other bugs in the development pipeline. Other reasonable benchmarks include a 'majority-class' model that predicts the most common output in the training set and a 'by-hand' model which invites a human to consider the inputs and generate an output.

The 'by-hand' model is often called a Wizard-of-Oz, or human-assisted model, and has been used at scale by large tech companies to help understand human interactions \cite{facebookM}. Similar to the \textit{whiteboard model}, these baselines will help to define essential features in your data for the given performance metric.

\paragraph{Step 6: Evaluate model} Your evaluation data will serve as a consistent comparison for model improvement. Test your model on your evaluation data segment. Track your key metric. The performance of your baseline model starts as your 'best,' and 'worst,' performing model. Keep your model performances as comparators as you iterate in Step 8: Re-evaluate and Iterate. What the limitations of your evaluation scheme? What are the unaccounted costs or errors? How does the model perform on the evaluation data and the safety suite designed in Step 3: Define the goal.

With each model evaluation iteration, it is essential to think about biases, fairness, and equality across diverse groups. Each iteration is a crucial checkpoint to communicate with stakeholders. Your stakeholders' discussions should include how they feel your model has addressed the ideas, interests, design requirements, concerns, and questions brought up in Step 2: Loop in Humans. How do they evaluate system performance? How would the baseline model impact them? Consider your stakeholders might be satisfied with your baseline model.

\paragraph{Step 7: Analyze trade-offs} You will make trade-offs as you make iterate models, and model parameters, towards optimizing your crucial metric. Consider these trade-offs by listing each of them and their associated impacts independently. Trade-offs often include factors such as cost, storage, learning speed, inference speed, computation complexity, model serving, deployment, and human interpretability. It helps to perform ablation studies which systematically remove model components to determine their relative contributions. Considering each of these trade-offs will help you iterate on your model development.

\paragraph{Step 8: Re-evaluate and iterate} Given the trade-offs defined in Step 7, review your key metrics. Ensure you capture all the information required before continuing. For instance, how do you log experimental parameters (i.e. model information) and results? Once you are confident that you can systematically make model improvements towards your evaluation metric, then it is time to iterate through Steps 5, 6, and 7. Once your evaluation performance converges, only then should you test your model(s) on the held out test set data segment. This testing should be used to compare models, and not to tune model parameters.

\section{Disseminate}
\label{sec:diss}

\paragraph{Step 9: Deploy the system} Present and test the system with your stakeholders and individuals you have not engaged with up to this point. When testing with humans, focus on usability. How are stakeholders interacting with your model? Usability can have a profound effect on the perceived quality and capability of models. These are valuable interactions, note how these humans interpret the performance of your system.

Consider that many humans may act against the system, by accident or on purpose. How will you handle attacks on your model? What are the technical security implications of model security and the value-based human-experience design choices you have made which may influence human-behaviour? What are the fail-safes and procedural safeguards and how can you adapt them during deployment? How are you communicating the risks of interaction?

\paragraph{Step 10: Communicate} The purpose of communication is to convey the key ideas to your audience clearly so that they may comprehend them with minimal effort. You should be able to state your key results and how it aligns with your hypothesis? Do your results match or contradict similar work? What are the limitations of the current model? How might these problems be addressed in the future? Do the results challenge any of the ideas or beliefs of the stakeholders?

When communicating the project, it is helpful to follow the ML Reproducibility checklist \cite{pineau2018repro}. Can you open source your code, data, models, and deployment? Consider how and why others might attempt reproduction.

\section{Conclusions}
\label{sec:conc}

Human-centred thinking for design and development can lead to substantial improvements in the development and adoption processes.  By empathizing with those invested in, impacted by, or adversaries of ML systems, developers can better serve the needs of all humans involved. Enabling human to efficiently and effectively interact with systems continues to be a key design challenge \cite{dudley2018review}. Human-centric design in ML can help address ongoing challenges of bias and unfairness and potentially improve the transparency and accountability of the choices which go into designing, developing and deploying new systems.

\printbibliography

\end{document}